\documentclass[10pt]{article}
\usepackage[OE]{express}

\begin{document}

\title{Photon statistics and bunching of a chaotic semiconductor laser}

\author{Yanqiang Guo,\authormark{1,2} Chunsheng Peng,\authormark{1,2} Yulin Ji,\authormark{1,2} Pu Li,\authormark{1,2} Yuanyuan Guo,\authormark{1,2} and Xiaomin Guo\authormark{1,2,*}}

\address{\authormark{1}Key Laboratory of Advanced Transducers and Intelligent Control System, Ministry of Education and Shanxi Province, Taiyuan University of Technology, Taiyuan 030024, China\\
\authormark{2}Institute of Optoelectronic Engineering, College of Physics and Optoelectronics, Taiyuan University of Technology, Taiyuan 030024, China}

\email{\authormark{*}guoxiaomin@tyut.edu.cn} %% email address is required



\begin{abstract}
The photon statistics and bunching of a semiconductor laser with external optical feedback are investigated experimentally and theoretically. In a chaotic regime, the photon number distribution is measured and undergoes a transition from Bose-Einstein distribution to Poisson distribution with increasing the mean photon number. The second order degree of coherence decreases gradually from 2 to 1. Based on Hanbury Brown-Twiss scheme, pronounced photon bunching is observed experimentally for various injection currents and feedback strengths, which indicates the randomness of the associated emission light. Near-threshold injection currents and strong feedback strengths modify exactly the laser performance to be more bunched. The macroscopic chaotic dynamics is confirmed simultaneously by high-speed analog detection. The theoretical results qualitatively agree with the experimental results. It is potentially useful to extract randomness and achieve desired entropy source for random number generator and imaging science by quantifying the control parameters.
\end{abstract}

 % REPLACE WITH CORRECT OCIS CODES FOR YOUR ARTICLE, MINIMUM OF TWO; Avoid using the OCIS codes for ¡°General¡± or ¡°General science¡± whenever possible.
%For a complete list of OCIS codes, visit: https://www.osapublishing.org/oe/submit/ocis/

%%%%%%%%%%%%%%%%%%%%%%% References %%%%%%%%%%%%%%%%%%%%%%%%%

%%%%%%%%%%%%%%%%%%%%%%%%%%  body  %%%%%%%%%%%%%%%%%%%%%%%%%%
\section{Introduction}

Semiconductor lasers (SCL) with external optical feedback (EOF) provide a rich platform to study the nonlinear effects and complex photonics dynamics \cite{Soriano13,Sciamanna15}. EOF is the most prominent configuration for achieving laser chaos. A common aspect of such systems is the presence of chaotic behaviour that consists of very short and random spiking of the laser intensity. The chaotic behaviour in these systems is extremely sensitive to time-delayed returning field of their own emission and was identified as early as the late 1960s \cite{Broom69}. To verify the observed chaotic dynamics, the Lang-Kobayashi (LK) model \cite{Lang80} is employed to describe the system qualitatively, and even quantitatively in some cases \cite{Torcini06}. With the growing understanding of the chaotic dynamics and its control, the optical-feedback-induced laser chaos has gained more and more attention for chaos-based secure communications \cite{Argyris05}, chaos key distribution \cite{Yoshimura12}, physical random number generation (RNG) \cite{Uchida08,Reidler09,Wang13}, chaotic optical sensing \cite{Lin04,Wang08}, and bioinspired information processing \cite{Brunner13}. Thus, it is indispensable to understand and elucidate the underlying mechanisms, the nonlinear dynamics under different conditions, and how to distinguish and identify different chaotic properties. Intensity statistics and autocorrelation (AC) associated to the dynamics of SCL with EOF are important and useful to characterize chaotic processes. The intensity statistics is closely related to the maximum extractible rate of randomness \cite{Reidler09,Argyris10,Oliver11} and post-processing techniques \cite{Kanter10} in RNG application. In optical chaos communications, the AC can be used as a good indicator of a bandwidth of chaotic laser limiting the speed of modulating chaotic carrier \cite{Locquet10}. Additionally, the AC function can present useful information about the weak-strong chaos transition in the laser system with EOF \cite{Heiligenthal11,Toomey14,Porte14}. Previous research has been mainly devoted to elucidating the intensity statistics of chaotic laser that operates in the low-frequency fluctuations (LFF) regime \cite{Heil99,Sukow99}. A typical characteristic of LFF is a sudden power drop followed by a gradual power recovery and the frequency of LFF is distinctly low compared to the laser's intrinsic relaxation oscillation frequency \cite{Sacher89}. This phenomenon occurs near the laser threshold for low-to-moderate feedback and has been explained as chaotic itinerancy with a drift \cite{Sano94,Fischer96}. The chaotic itinerancy predicts fast chaotic pulsations of the output intensity. Since the pulsing behavior skews the intensity distribution, the probability density distributions of the laser intensity becomes extremely asymmetric \cite{Sukow99}. As the pump current and feedback strength are increased, the laser linewidth is broadened greatly from a few MHz to tens of GHz and the LFF typically evolves into the laser's coherence collapse (CC) \cite{Sacher89}. The research on the intensity statistics of high-dimensional chaotic waveforms in the CC regime is sparse, in contrast to that of LFF dynamics \cite{Li14}. However, it is still unclear whether the intensity statistics of chaotic laser would change when varying the injection current and feedback strength. Currently, there is a significant discrepancy between experimental and theoretical probability density distributions \cite{Li14}. Recent researches reveal that the photon statistics is a candidate to indicate random-intensity fluctuations associated with the emission of chaotic laser \cite{Albert11}. Photon number distribution and photon correlations are more sensitive to control parameters compared to probability density distributions of laser intensity \cite{Schulze14}. Photon statistical investigation of chaotic laser is also rare in the continuous transition region between the LFF regime and the fully developed CC chaotic regime.

 The photon statistics and correlations are pivotal and fundamental in characterizing the quantum statistics of light sources. The pioneering experiments conducted by Hanbury Brown and Twiss (HBT) were a landmark linking the temporal and spatial second-order degree of coherence g$^{(2)}$($\tau ,x$) of a thermal source \cite{Hanbury56}. The g$^{(2)}$($\tau ,x$) was formalized by Glauber in the 1960s \cite{Glauber63} and described by correlation function within statistical optics \cite{Mandel95}, which has information on both the nature and dynamics of the photon emission process of the underlying field. Subsequently, the observation of g$^{\text{(2)}}$ effect is harnessed in numerous experiments and applications, such as measuring photon bunching and photon number distribution \cite{Arecchi65}, characterizing nonclassicality of light field \cite{Kimble77}, and so on. With the rapid development of photon counting technology, single photon counting detection, as the most sensitive and very widespread method of optical measurement, has been expanded towards quantum information \cite{Hadfield09} and precision measurement applications \cite{Banaszek09}. Up until now, g$^{(2)}$($\tau $) already allows one to categorize different light fields and carry a great deal of information on the photon statistical properties of a light field. In general, g$^{(2)}$($0$) (i.e., at zero time difference) $>1$ or Bose-Einstein distribution, characteristic of an incoherent or chaotic light, while $g^{(2)}(0)=1$ or Poissonian distribution, characteristic of a coherent light or a classical stable field, and $g^{(2)}(0)<1$ or sub-Poissonian distribution, characteristic of a quantum light emission \cite{Davidovich96}. In this regard, $g^{(2)}(\tau )$ is fundamentally different from the first-order degree of coherence, which serves as a description of phase and cannot sufficiently extract quantum properties of light field. Thus, HBT scheme combined with single-photon detection have boosted the study of spatial interference \cite{Schultheiss16}, ghost imaging \cite{Ryczkowski16}, azimuthal HBT effect \cite{Loaiza16}, deterministic manipulation and detection of single-photon source \cite{Guo12}, etc. Moreover, the random or chaotic properties of light is an essential element of the HBT effect and have been applied in full-field imaging science \cite{Redding12} and nanophotonics \cite{Hayenga16}. However, quantum statistics of chaotic laser is unexplored. More recent experiments of photon statistics in chaotic regime aim to bridge the gap between chaotic laser and quantum optics \cite{Albert11,Hagerstrom15}. A comprehensive understanding of photon statistics and correlation of chaotic laser remains an open question.

In this paper, we present the experimental and numerical results of photon statistics and coherence of chaotic laser based on single photon counting technique. The chaotic laser operating in the mW (high-gain) regime, consists of SCL with time-delayed optical feedback. In order to compare macroscopic dynamics and photon statistical transition, we simultaneously measure the optical signals by photon-counting method and high-speed analog detection. Implementing HBT interferometry, we demonstrate that the control parameters of chaotic laser, i.e., feedback strength and injection current, can substantially affect the photon correlation. The photon number distribution of the chaotic laser undergoes a transition from Bose-Einstein distribution to Poissonian distribution as the mean photon number increases. It shows a good agreement between experimental photon number distribution and theoretical fitting without post-processing. In the chaotic regime, the laser emission shows bunching effect, and the control parameters modify exactly the laser performance to be more bunched. Lower injection current and stronger feedback strength are beneficial for enhancing bunching effect. The macroscopic chaotic dynamics is confirmed experimentally. The phenomenon is also verified numerically using the LK equations where the results are in good agreement with the experimental data. This demonstration well reveals photon statistics and coherence of chaotic laser and provide a necessary and better understanding of chaotic process. In this sense, it contributes to the research of chaos with quantum optics.

\section{Experimental setup and Theoretical Model}

The experimental setup is schematically illustrated in Fig. 1. The chaotic laser is composed of a distributed feedback laser diode (DFB-LD, WTD LDM5S752) subject to external optical feedback, operating at 1.55 $\mu $m and with threshold current J$_{th}$ of 10.5 mA. The temperature and current of DFB-LD are stabilized to an accuracy of 0.01 K and 0.1 mA, respectively. The DFB-LD connects to an 80:20 optical coupler (OC1) whose principal output passes through a variable optical attenuator (VOA1) onto a optical circulator, forming a fiber-based feedback cavity with a time delay of 99.85 ns. The VOA1 and polarization controller (PC) are used to accurately define the feedback conditions. The optical feedback drives the DFB-LD into high-frequency chaotic oscillation. The output of chaotic laser is split via 50:50 optical coupler (OC2). At one port photon statistics and coherence are measured based on a HBT scheme: the laser emission is coupled out and collimated by triplet lens coupler. The freespace beam goes through a filter and is divided intensity-equally into two parts. The transmitted and reflected photons eventually arrive at dual-channel single photon detector (DSPD, Aurea Technology LYNXEA-NIR-M2-SM-01). The quantum efficiency of DSPD is 25\% at 1550 nm and the counting rate is controlled below 0.4 Mcounts/s by a variable optical attenuator (VOA2). The DSPD outputs are then connected to a time-correlated coincidence unit and the time difference between the two signals originating from two emitted photons is repeatedly measured with 60 ps resolution. The other port of OC2 is used for detection of intensity and frequency dynamics by two 50 GHz photodetectors (PD, Finisar XPDV2120RA). All data are recorded simultaneously using a 26.5 GHz RF spectrum analyzer (SA, Agilent N9020A, 3 MHz RBW, 3 KHz VBW) and a 40 Gsamples/s real-time oscilloscope (OSC, Lecroy, LabMaster10-36Zi) with 36 GHz bandwidth.

\begin{figure}[ht!]
\centering\includegraphics[width=11cm]{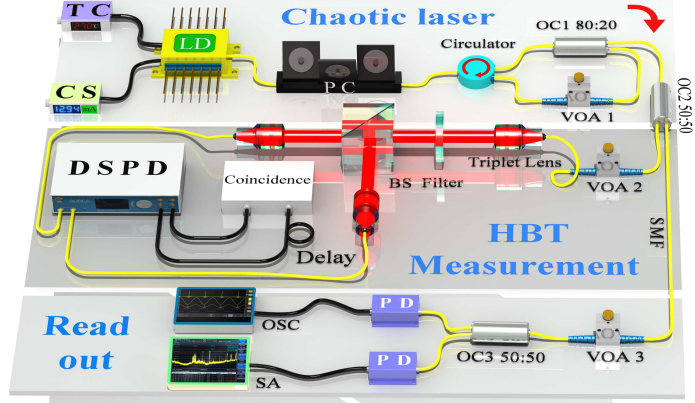}
\caption{Schematic illustration of the experimental setup. LD: distributed feedback laser diode; PC: polarization controller; TC: temperature controller; CS: current source; OC1, OC2 and OC3: optical coupler; VOA1, VOA2 and VOA3: variable optical attenuator; SMF: single mode fiber; BS: beam splitter; DSPD: dual-channel single photon detector; PD: Photodetector; OSC: oscilloscope; SA: spectrum analyzer.}
\end{figure}

In the high-gain regime, chaotic dynamics of a single-mode semiconductor laser with time delayed optical feedback can be well modeled by the LK equations, which are capable of confirming and reproducing the observations \cite{Lang80}. The LK equations for the three real variables of the optical field amplitude or the photon flux density $E(t)$, the field phase $\varphi (t)$, and the carrier density $N(t)$ are described as follows:
\begin{equation}
\dot{E}(t)=\frac{1}{2}[G(t)-\tau _{p}^{-1})E(t)+\kappa E(t-\tau _{ext})\cos
[\phi (t)],
\end{equation}%
\begin{equation}
\dot{\varphi}(t)=\frac{\alpha }{2}[G(t)-\tau _{p}^{-1}]-\kappa E(t-\tau
_{ext})E(t)^{-1}\sin [\phi (t)],
\end{equation}%
\begin{equation}
\dot{N}(t)=\frac{J}{e}-\frac{N(t)}{\tau _{N}}-G(t)\left\vert E(t)\right\vert
^{2},
\end{equation}%
\begin{equation}
\phi (t)=\omega \tau +\varphi (t)-\varphi (t-\tau _{ext}),
\end{equation}
where $G(t)=G_{N}[N(t)-N_{0}]/(1+\varepsilon \left\vert E(t)\right\vert ^{2})$ is the nonlinear optical gain (with $G_{N}$ being the gain coefficient and $\varepsilon $ the saturation coefficient), $N_{0}$ is the carrier density at transparency, $\kappa =(1-r_{in}^{2})r_{0}/(r_{in}\tau _{in})$ is the optical feedback strength, $r_{in}$ is the reflectivity of the internal cavity, $r_{0}$ is the reflectivity of the external mirror, $\tau _{in}$ is the optical round-trip time in internal cavity, $\tau _{p}$ is the photon lifetime, $\tau _{N}$ is the carrier lifetime, $\alpha $ is the linewidth-enhancement factor, $\tau _{ext}$ is the feedback delay time of the external cavity, $\omega =2\pi c/\lambda $ is the angular optical frequency, $c$ is the speed of light, and $\lambda $ is the optical wavelength, and $J=\rho J_{th}$ is the injection current density (with $\rho $ being the pump factor). The laser threshold current results from Eq. (3) as $J_{th}=(1/\tau_{N})[N_{0}+1/(G_{N}\tau _{p})]\approx 10.5$ $mA$, corresponding to $J_{th}$ of the DFB laser. The following values for the above parameters are fixed at $\alpha =5$, $\tau _{p}=2.5$ $ps$, $\tau _{N}=2.3$ $ns$, $G_{N}=2.56\times 10^{-8}$ $ps^{-1}$, $N_{0}=1.35\times 10^{8}$, $\tau _{ext}=99.85$ $ns$, $\lambda =1.55$ $\mu m$, $\varepsilon =5\times 10^{-7}$. We numerically integrated the Eqs. (1)--(3) with a fourth order Runge-Kutta routine by employing time steps of $h=2\times 10^{-12}$ $s$.

For the initial conditions, Eqs. (1)--(3) are solved many times (typically $10^{5}$), each solution being referred to as a \textquotedblleft realization\textquotedblright\ of the laser operation. The second-order degree of coherence $g^{(2)}(\tau )$ of optical field can be calculated as
\begin{equation}
g^{(2)}(\tau )=\frac{\left\langle E^{\ast }(t)E^{\ast }(t+\tau )E(t+\tau
)E(t)\right\rangle }{\left\langle E^{\ast }(t)E(t)\right\rangle ^{2}}=\frac{%
\int_{t}I(t)I(t+\tau )dt}{(\int_{t}I(t)dt)^{2}}=\frac{\left\langle
n_{ph}(t)n_{ph}(t+\tau )\right\rangle }{\left\langle n_{ph}(t)\right\rangle
^{2}},
\end{equation}
where $I(t)=\left\vert E^{\ast }(t)E(t)\right\vert $ denotes the laser intensity, $\tau $ is the time difference between the two photon detection events, $n_{ph}(t)$ is the photon number and $\left\langle \cdot \right\rangle $ designates the statistical averaging that is done over a large ensemble of different realizations of the laser field. For coherent light, the photon number distribution $P(n_{ph})$ obeys Poisson distribution $P(n_{ph})=\left\langle n_{ph}\right\rangle ^{n_{ph}}e^{-\left\langle n_{ph}\right\rangle }/n_{ph}!$, where $\left\langle n_{ph}\right\rangle =\sum_{n_{ph}}n_{ph}P(n_{ph})$ is the mean photon number. For chaotic light, the photon number distribution $P(n_{ph})$ obeys Bose-Einstein distribution $P(n_{ph})=\left\langle n_{ph}\right\rangle ^{n_{ph}}/(1+\left\langle n_{ph}\right\rangle )^{n_{ph}+1}$. Figure 2(a) shows the calculated emission intensity for $J=1.5J_{th}$, $\kappa =35$ $ns^{-1}$, $\kappa =50$ $ns^{-1}$, $\kappa =65$ $ns^{-1}$, and the parameter values mentioned above. For $\kappa =35$ $ns^{-1}$, $\kappa =50$ $ns^{-1}$, and $\kappa =65$ $ns^{-1}$ the model predicts chaotic intensity fluctuations. The related photon number distributions are depicted in Fig. 2(b). The histograms from the LK model are better fitted by Bose-Einstein distribution (blue solid curve), whereas the Poisson distribution (black dotted curve) shows relatively poor fit. The chaotic intensity fluctuations and the spiked emission of photons result in bunching effect with $g^{(2)}(0)>1$. As the optical feedback strength $\kappa $ increases the bunching effect becomes strong, and the associated $g^{(2)}(\tau )$ results are shown in Fig. 2(c).
\begin{figure}[ht!]
\centering\includegraphics[width=10cm]{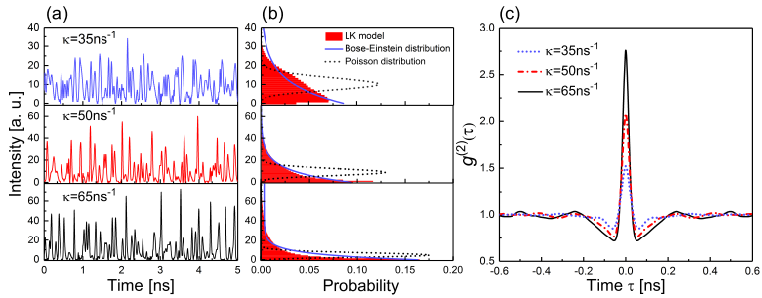}
\caption{Intensity time traces, photon number distribution and second-order degree of coherence for $J=1.5J_{th}$ with $\kappa =35$ $ns^{-1}$, $\kappa =50$ $ns^{-1}$, and $\kappa =65$ $ns^{-1}$. (a) Calculated emission intensity and (b) associated photon number distribution of a semiconductor laser with external optical feedback. The blue solid and black dotted curves are the Bose-Einstein and Poisson distributions. (c) The corresponding $g^{(2)}(\tau )$ are obtained from emission intensity for various $\kappa =35$ $ns^{-1}$ (blue dotted curve), $\kappa =50$ $ns^{-1}$ (red dash-dotted curve), and $\kappa =65$ $ns^{-1}$ (black solid curve).}
\end{figure}
For $\kappa =35$ $ns^{-1}$, the central bunching maximum is observed with $1<g^{(2)}(0)<2$ that is associated with sub-chaotic light. For $\kappa =50$ $ns^{-1}$, the bunching maximum is obtained with a magnitude $g^{(2)}(0)\sim 2$, which indicates that the light approaches fully chaotic limit ($g^{(2)}=2$), or chaotic limit. For strong optical feedback strength [$\kappa =65$ $ns^{-1}$ in Fig. 2(c)], the pronounced bunching maximum is observed with a magnitude $g^{(2)}(0)$ substantially exceeding 2 and the light becomes super chaotic, which is more chaotic than the fully chaotic light mentioned above. Owing to the laser's intrinsic relaxation oscillation, slight oscillations of $g^{(2)}(\tau )$ appear around the central bunching area.

\section{Results}

We start by evaluating the photon statistics of chaotic laser. Figures 3(a) and 3(b) show the measured time traces of the chaotic laser and its associated RF spectrum operating at $J=1.5J_{th}$ and experimental feedback strength $\eta =12.8\%$. In our experiment, $\eta =25\%$ of the optical power is fed back for $\kappa =20$ $ns^{-1}$ and we keep it below. Comparing with the noise floor, the output of the laser shows high bandwidth and large amplitude fluctuation, which indicate that the laser is operating in a chaotic region. In Fig. 3(a), the gray line (red for online version) is noise floor from the stable operation laser. According to the $80\%$ bandwidth definition, the bandwidth of the chaotic laser is about 9.85 GHz, as shown in Fig. 3(b).

\begin{figure}[ht!]
\centering\includegraphics[width=10cm]{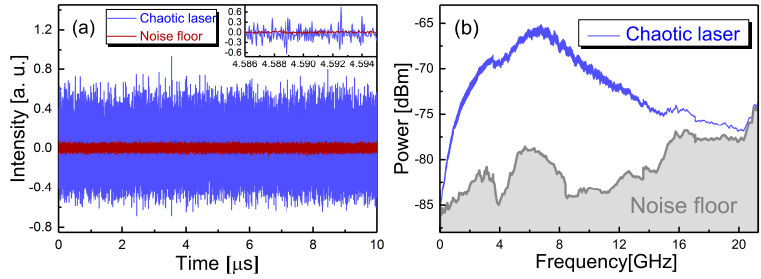}
\caption{Characteristics of chaotic laser when $J=1.5J_{th}$ and $\eta =12.8\%$. (a) Measured time traces and (b) power spectrum of the chaotic laser. The inset in (a) is the time series in a shorter time interval.}
\end{figure}

Under this condition, the measured photon number distribution $P(n_{ph})$ and second-order degree of coherence $g^{(2)}(0)$ of chaotic laser are illustrated in Fig. 4. The incident photon number is controlled by finely adjusting the programmable VOA2. With an increase of the mean photon number, the photon number distribution changes continuously from Bose-Einstein distribution to Poisson distribution and the $g^{(2)}(0)$ decreases gradually from 2 to 1. The maximum value of measured $g^{(2)}(0)$ is 2.02 and the associated photon number distribution is almost identical to the Bose-Einstein distribution of the same mean photon number ($\left\langle n_{ph}\right\rangle =0.69$). As we increase the mean photon number, the photon number distribution of chaotic laser gradually deviates from the Bose-Einstein distribution. When the mean photon number $\left\langle n_{ph}\right\rangle $ reaches 1.8, the measured photon number distribution is between the Bose-Einstein distribution and Poisson distribution and the $g^{(2)}(0)$ becomes 1.21. As the mean photon number is increased further, the photon number distribution gets closer to the Poisson distribution and the $g^{(2)}(0)$ is reduced. Eventually, when the mean photon number  $\left\langle n_{ph}\right\rangle $ is 2.61, the photon number distribution is nearly identical to the Poisson distribution and the corresponding $g^{(2)}(0)$ is 1.03. Due to the fact that the multiphoton events increase, the photon arriving interval becomes shorter compared to the sampling time. From another point of view, the sampling time gets longer than the coherence time of the emitted photons. Thus, the measured photon number distribution gradually approaches Poisson distribution. Without any post-processing, the experimental results are in good agreement with the theoretical fitting. The accuracy is improved compared to the measurement of macroscopic intensity statistics \cite{Li14}. The measured photon statistics of the light field also can be profoundly affected by experimental conditions. In this work, we choose the proper experimental conditions (e.g., below 1.62 for the mean photon number and 80 Counts/s for the photon count rate) and the results can reflect correctly the photon statistics of light field \cite{Li07}.

\begin{figure}[ht!]
\centering\includegraphics[width=8.5cm]{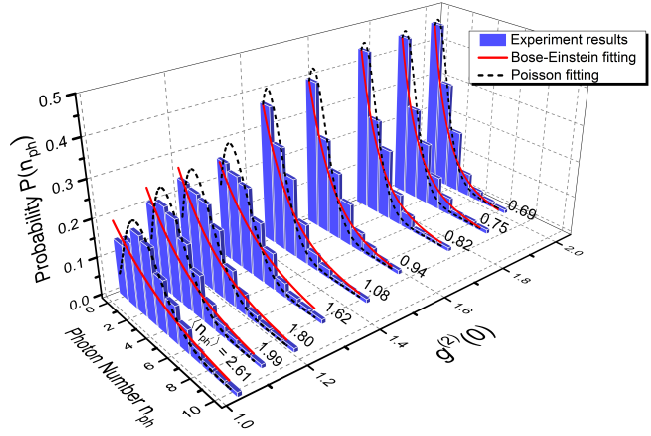}
\caption{The blue histograms are the measured photon number distribution $P(n_{ph})$, corresponding to $g^{(2)}(0)$ for the chaotic laser with different mean photon numbers $\left\langle n_{ph}\right\rangle $ at the input. The red solid and black dashed curves are the Bose-Einstein and Poisson fitting, respectively.}
\end{figure}

Despite its utility, the photon number distribution $P(n_{ph})$ is a single description and does not constitute a complete representation of the light. A unique description requires knowledge of the photon correlation $g^{(2)}$, which we have shown in Fig. 5 for various injection current $J$ and feedback strength $\kappa $ in the chaotic regime. The $g^{(2)}(0)$ also characterizes the randomness of the associated photon number $n_{ph}$ and is a measurement of photon bunching, which refers to the tendency of photons to arrive together. As the $g^{(2)}(0)$ increases, the photons become more bunched and chaotic \cite{Kondakci15}. Figure 5 shows theoretical 5(a) and experimental 5(b) $g^{(2)}(0)$ as functions of $\kappa (\eta )$ for five different values of $J$ in chaotic regime: $1.07J_{th}$, $1.12J_{th}$, $1.2J_{th}$, $1.5J_{th}$, and $2.0J_{th}$. At all the set $J$ values, the $g^{(2)}(0)$ gradually grows as $\kappa (\eta )$ increases. The $\kappa (\eta )$ affects little on the threshold $J_{th}$, and there is a negligible influence on photon correlation measurement. The experimental results shows good agreement with the theory. Additionally, the bunching effect become weak as the inject current $J$ increases. In Figs. 5(c) and 5(d), the theory and experiment for $g^{(2)}(0)$ at various $J/J_{th}$ and $\kappa (\eta )$ are shown. At all $\kappa (\eta )$ shown, the $g^{(2)}(0)$ decreases monotonically as the current $J$ increases and the theory is in good agreement with the experiment. For larger $J$, more multiphoton events result in the $g^{(2)}(0)$ falls faster, leading to approach the stable laser output. Near-threshold injection currents and strong feedback strengths are beneficial for enhancing bunching effect. The pronounced photon bunching is helpful to boost the image contrast in ghost imaging. Moreover, the bunched chaotic laser with low coherence is well suited for speckle-free full-field imaging \cite{Redding12}. The results also provide a basic picture and understanding of the higher order correlation of chaotic light in the single photon counting regime.

\begin{figure}[ht!]
\centering\includegraphics[width=9.5cm]{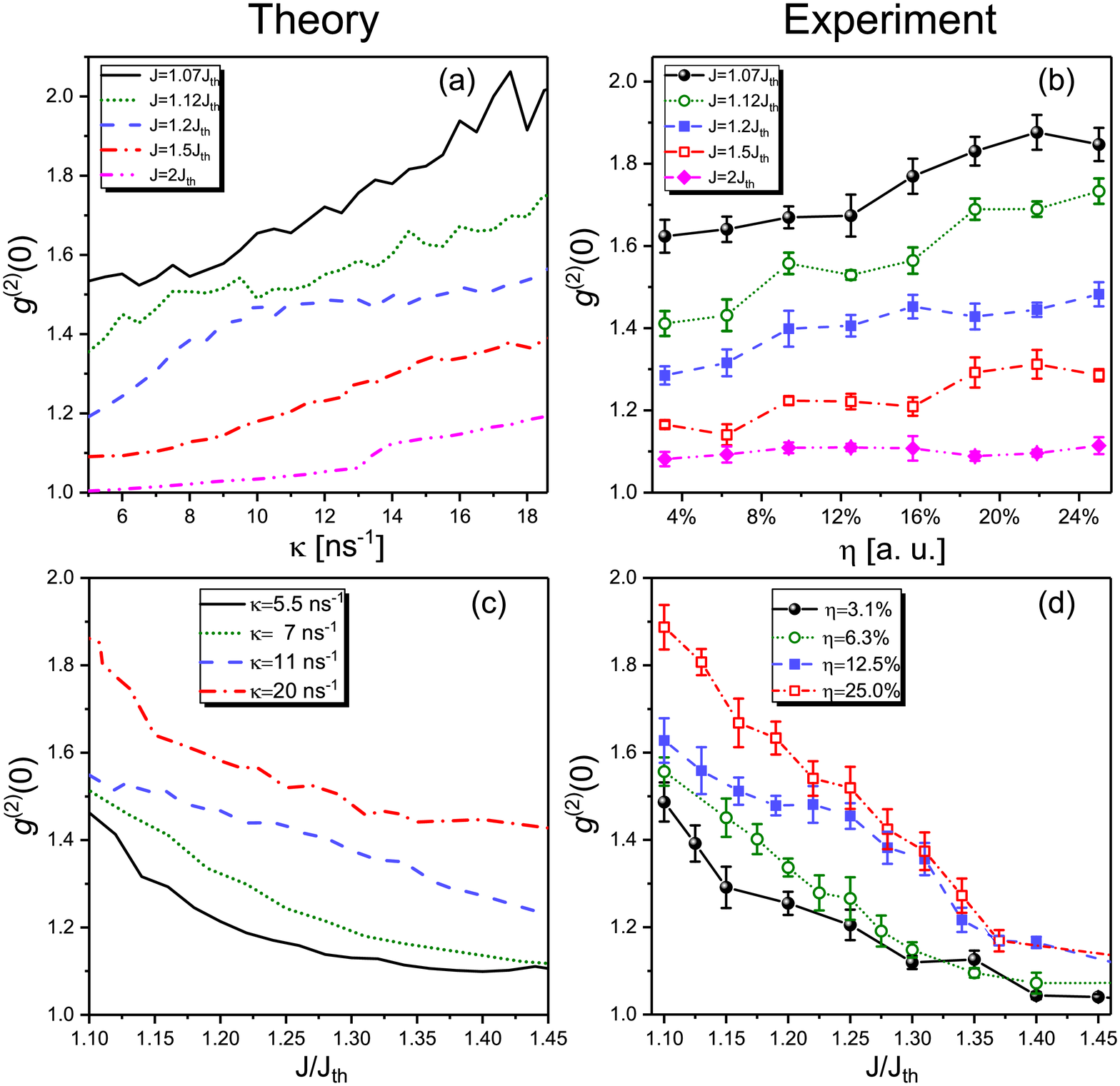}
\caption{(a), (c) Theoretical and (b), (d) experimental results for $g^{(2)}(0)$ at five $J$ (a), (b): $1.07J_{th}$, $1.12J_{th}$, $1.2J_{th}$, $1.5J_{th}$, and $2.0J_{th}$; four $\kappa (\eta )$ (c), (d): $5.5$ $ns^{-1}$ $(3.1\%)$, $7$ $ns^{-1}$ $(6.3\%)$, $11$ $ns^{-1}$ $(12.5\%)$, $20$ $ns^{-1}$ $(25\%)$.}
\end{figure}

In what follows, we confirm the chaotic dynamics to help interpret the near-threshold bunching effect. Figure 6 depicts the measured temporal waveform and RF spectrum for three values of $\eta =3.1\%$, $\eta =12.5\%$, and $\eta =25\%$, when $J=1.07J_{th}$ [Figs. 6(a1)-6(a4)], $J=1.2J_{th}$ [Figs. 6(b1)-6(b4)], and $J=1.5J_{th}$ [Figs. 6(c1)-6(c4)]. At all the three $J$ values, the amplitude fluctuation and bandwidth of chaotic laser increase as the feedback strength $\eta $ grows. From Fig. 6(a4) to Fig. 6(c4), the bandwidth of the laser is broadened from 2.65 GHz to 9.9 GHz as the injection current and feedback strength increase. In the near-threshold regime, wider bandwidth and larger amplitude fluctuation contribute to obtaining stronger bunching effect when the injection current is fixed. While the feedback strength is fixed, it's an opposite case that stronger bunching effect happens on the bandwidth and amplitude fluctuation drop.

\begin{figure}[ht!]
\centering\includegraphics[width=11cm]{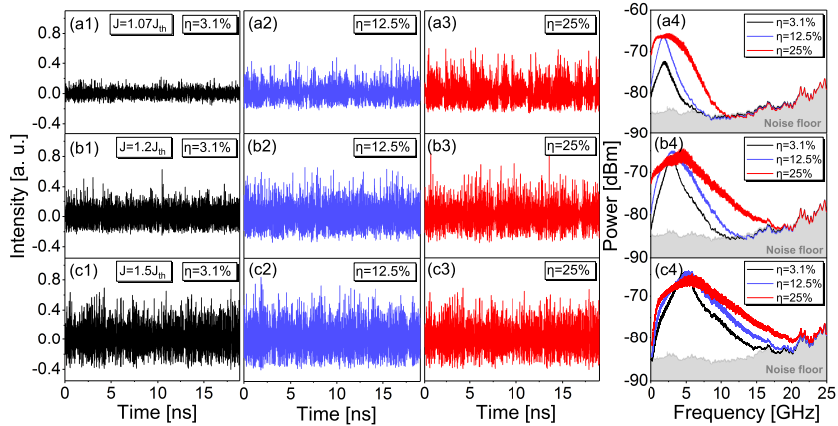}
\caption{Measured temporal waveform and RF spectrum of chaotic signals for three $\eta $, when (a) $J=1.07J_{th}$, (b) $J=1.2J_{th}$, and (c) $J=1.5J_{th}$.}
\end{figure}

In addition, the external-cavity delay provide useful information on the chaotic dynamics and indicate the weak-strong chaos transition \cite{Porte14,Li14}. The maximum height $h$ of normalized autocorrelation function near the first delay echo is inversely proportional to the chaotic strength. Figure 7 illustrates theory 7(a) and experiment 7(b) for $h$ at various $J$ and $\kappa (\eta )$. At all $J$ shown, the $h$ follows a nonmonotonic dependence on the feedback strength $\kappa (\eta )$, revealing a dip for intermediate $\kappa (\eta )$. The theory is in good agreement with experiment. In the week feedback regime, the chaotic strength and bunching effect are both enhanced as $\kappa (\eta )$ increases. On the other side (strong feedback), the chaotic laser becomes more bunched but the chaotic strength decreases when increasing $\kappa (\eta )$.

\begin{figure}[ht!]
\centering\includegraphics[width=9cm]{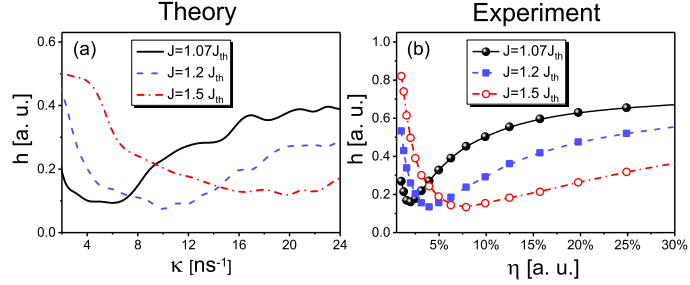}
\caption{(a) Theoretical and (b) experimental results for $h$ at various $\kappa (\eta )$ and three $J$: $1.07J_{th}$, $1.2J_{th}$, and $1.5J_{th}$.}
\end{figure}

\section{Conclusions}

To conclude, the photon statistics and second-order photon correlation $g^{(2)}$ are investigated theoretically and experimentally for a SCL with optical feedback and current modulation in the chaotic regime. The photon number distribution of the laser operating in a chaotic regime transitions continuously from Bose-Einstein distribution to Poisson distribution as the mean photon number increases. The theoretical results predict the pronounced photon bunching which is observed in HBT experiment. We have found a good agreement between the theoretical model and experimental observations for the $g^{(2)}(0)$ as functions of $J$ and $\kappa (\eta )$. To describe the photon statistics and bunching of SCLs under optical feedback in the high-gain regime, the LK model is well employed. The bunching effect is enhanced by increasing the feedback strength near the threshold current. The results also provide a better understanding of the super-chaotic bunching phenomena. The new type of photon bunching spectroscopy with the current and optical feedback detuning allow us to extract information about photon statistics and randomness. In this sense, the finding will boost the study of ghost imaging utilizing chaotic laser with tunable photon correlation and speckle-free full-field imaging with low-coherence chaotic laser, and contribute to improvements of chaos-based random number generation and secure communication with chaos synchronization.

\section*{Funding}

National Natural Science Foundation of China (NSFC) (61405138, 61505136, 61775158, 61671316); Shanxi Scholarship Council of China (SXSCC) (2017-040); Natural Science Foundation of Shanxi Province (201701D221116).

\section*{Acknowledgments}

The authors thank H. Shen for helpful discussions.

\end{document}